\documentclass[a4paper,12pt]{scrartcl}

\usepackage[]{graphicx}\usepackage[]{color}

\usepackage{alltt}
\usepackage[left= 2.5cm,right = 2.5cm, bottom = 3.0cm,top=2cm]{geometry}
\usepackage[latin1]{inputenc} 
\usepackage{setspace}
    \usepackage[flushleft]{threeparttable}

\usepackage{graphicx,psfrag,epsf}

\usepackage{amsmath}
\usepackage{amsfonts}
\usepackage{amssymb}
\usepackage{graphics}
\usepackage{graphicx}
\usepackage{epstopdf}
\usepackage[nottoc]{tocbibind}
\usepackage[authoryear]{natbib}
\newcommand{\GG}[1]{}

\newlength{\minuslength}
\usepackage{subfig}

\usepackage{bm}

\usepackage{booktabs}

\usepackage{acronym} 
\usepackage{here} 

\usepackage{titling}

\title{A large non-Gaussian structural VAR with application to Monetary Policy\thanks{We thank Boris Blagov, Ralf Br\"uggemann, Robert Czudaj, Sascha Keweloh and Linus N\"using for their valuable feedback. Jan Pr\"user gratefully
acknowledges the support of the German Research Foundation (DFG, 468814087).}} 

\author{Jan Pr\"user$^{a}$
\\[0.2cm]
$^{a}${\small TU Dortmund\thanks{Fakult\"at Statistik, 44221 Dortmund, Germany, e-mail: \texttt{prueser@statistik.tu-dortmund.de\,}}} }

\begin{document}

\date{\today}
\maketitle
\begin{abstract}
	\begin{singlespace}
	\noindent
We propose a large structural VAR which is identified by higher moments without the need to impose economically motivated restrictions. The model scales well to higher dimensions, allowing the inclusion of a larger number of variables. We develop an efficient Gibbs sampler to estimate the model. We also present an estimator of the deviance information criterion to facilitate model comparison. Finally, we discuss how economically motivated restrictions can be added to the model. Experiments with artificial data show that the model possesses good estimation properties. Using real data we highlight the benefits of including more variables in the structural analysis. Specifically, we identify a monetary policy shock and provide empirical evidence that prices and economic output respond with a large delay to the monetary policy shock.
	\end{singlespace}
\end{abstract}
\bigskip
 
 
	 \bigskip

\thispagestyle{empty} 
\newpage


\pagenumbering{arabic}
\section{Introduction}

In econometrics, structural vector autoregressive (SVAR) models are key tools for analysing dynamic relationships among time series data, but identifying structural shocks remains a critical challenge. Traditional identification relies on economically motivated restrictions, such as short-run zero restrictions \citep{sims1980macroeconomics}, sign restrictions \citep{uhlig2005effects}, or proxies \citep{mertens2013dynamic}, to extract meaningful shocks. However, these restrictions cannot be formally tested, and second moments (covariances) alone are insufficient for full identification of structural parameters, see \cite{kilian2017structural}.

Recent advances have shown that higher order moments under non-Gaussian shocks, offer an alternative method of statistical identification, see \cite{lewis2024identification} for a review.\footnote{A related but distinct approach is to achieve identification through time-varying volatility, which also provides an additional source of information to distinguish structural shocks, see for example, \cite{rigobon2003identification}, \cite{lanne2010structural}, \cite{lutkepohl2020bayesian}, \cite{lewis2021identifying} and \cite{lutkepohl2024partial}.} By exploiting the additional information contained in higher unconditional moments, it is possible to achieve identification without the need for imposed economic restrictions. Non-Gaussian shocks allow for identification through the mutual independence of error terms rather than relying solely on second moments, see \cite{comon1994idependent}.  This result has been exploited in various ways, see e.g., \cite{lanne2017identification}, \cite{gourieroux2017statistical}, \cite{guay2021identification}, \cite{lanne2023identifying}, \cite{keweloh2021generalized}, \cite{braun2023importance} and \cite{hafner2024statistical}. However, these studies focus on SVAR models with a small number of variables. Increasing the number of variables raises concerns about overfitting as well as computational challenges. Computational challenges arises because these models are typically estimated using numerical maximisation algorithms (see e.g. \cite{keweloh2021generalized}) or Metropolis-Hastings-type algorithms (see e.g. \cite{lanne2023identifying}) and these algorithms may not scale well to higher dimensions.

Since the influential work of \cite{banbura2010large}, there has been growing interest in using large VARs with dozens or even hundreds of dependent variables for structural analysis and forecasting  (see, among others,
\cite{bloor2010analysing}, \cite{carriero2009forecasting}, \cite{carriero2012forecasting}, \cite{giannone2015prior}, \cite{jarocinski2017granger}, \cite{huber2019adaptive}, \cite{chan2024large} and \cite{hou2024large}).\footnote{Note that these studies do not focus on structural identification using higher moments.} This trend is partly motivated by the need to address problems arising from modelling too few variables, such as omitted variable bias, which can distort forecasting, policy advice, and structural analysis. By expanding the set of relevant variables, large VARs reduce concerns about informational deficiency, as pointed out in earlier work by \cite{hansen2019two} and \cite{lippi1993dynamic,lippi1994var}. These authors argue that when econometricians consider a narrower information set than economic agents, the model becomes non-fundamental, and structural shocks cannot be fully recovered. 
Large VARs also provide a practical solution to the challenge of mapping economic variables onto data that is often not unique. For example, inflation could be measured by different indices, such as the consumer price index or the gross domestic product deflator. Including multiple series for the same variable in the model helps to mitigate the arbitrary choice of data representation as argued by \cite{loria2022economic}. 
Large VARs are richly parameterised and prone to overfitting, but overfitting concerns can be effectively addressed using shrinkage techniques, such as the Minnesota prior see e.g. \cite{ingram1994supplanting} and \cite{cross2020macroeconomic}. 

In this paper, we propose a large non-Gaussian SVAR model with factor structure of the errors, which extends upon the existing literature that has focused on small non-Gaussian SVARs. Our approach introduces non-Gaussian shocks in a high-dimensional setting and aims to explore how higher moments and mutual independence of errors can be used to improve identification in large VAR environments, thereby overcoming limitations associated with both traditional economically motivated restrictions and small-dimensional models.

Recently it has become popular to assume that VAR errors have a factor structure. These factors are interpreted as structural shocks, see \cite{Korobilis2022} and \cite{chan2022large}. This has the advantage that when an additional variable is added to the VAR, it is not necessarily the case that an additional structural shock needs to be added. For example, if a researcher adds different measures of prices or economic activity she does not wish to add additional structural shocks to the model. Instead in SVARs with many variables it is reasonable to assume that the number of structural shocks may be much smaller than the number of variables.\footnote{A significant reduction in the number of shocks simplifies the process for researchers to accurately label them in a statistically identified SVAR, a topic we delve into further below.} From a computational point of view the factor structure allows for equation-by-equation estimation, which allows the model to scale to large dimensions as demonstrated by \cite{Korobilis2022}, \cite{chan2022large} and \cite{banbura2023drives}. \cite{Korobilis2022} suggests using sign restrictions on the factor loadings to archive set identification. \cite{chan2022large} combines the sign restrictions with stochastic volatility and \cite{banbura2023drives} combines the sign restrictions with a proxy variable to archive point identification.

We propose a large VAR model with non-Gaussian factors that scales well to higher dimensions. This model allows for the statistically identification of the structural shocks when they are both non-Gaussian and mutually independent without the need to impose any additional economically motivated restrictions. To address overfitting concerns in our richly parametrised model we use the Minnesota-type adaptive hierarchical prior suggested by \cite{chan2024large}. While the prior provides regularisation from a frequentist perspective it also has heavy tails to mitigate biases of large coefficients. We develop a Gibbs sampler that allows efficient sampling from the joint posterior distribution and scales well to higher dimensions. 
To compare different model specifications (e.g. models with different numbers of factors), we develop an estimator of the Deviance Information Criterion (DIC) proposed by \cite{spiegelhalter2002bayesian}. The DIC model can also be used to empirically assess the plausibility of over-identifing economic restrictions in our model framework as discussed next.

While our model uses information of higher moments to archive statistical identification without any further restrictions, it is still useful to consider how economically motivated restrictions can be added. Adding economic restrictions can serve two main purposes. First, we can incorporate economic restrictions to strengthen identification by higher moments, see \cite{carriero2024blended}.  In addition, identification based on economic prior knowledge offers natural solutions to the labelling problem, see \cite{braun2023importance}. Second, identification based on higher moments can be used to test economically motivated restrictions against the data. We can do this by using posterior summaries of the model parameters directly, or by comparing the DIC of the unrestricted model with the model in which we impose the economic restrictions. 


We demonstrate the usefulness of our approach using both artificial data and real data. Experiments with artificial data show the ability of our model to archive identification and provide reasonable estimates in finite samples. It has good frequentist estimation properties, providing unbiased estimates and credible bands with correct coverage rate. An empirical application demonstrates the advantages of our higher moments approach for structural identification in a high-dimensional setting. In particular, we use our model to identify a monetary policy shock. The model is estimated with the time series data used by \cite{uhlig2005effects}, enriched with additional measures of prices and economic activity as well as variables to capture information in financial and labour markets. While the structural shocks are identified statistically, a researcher still needs to attach an economic interpretation to them. We present different strategies for labelling the monetary policy shock, all of which lead to the same result. It turns out that prices and output respond with a large delay to the identified monetary policy shock. Finally, we extend our model with the proxy variable constructed by \cite{romer2004new} and provide empirical evidence against exogenous exclusion restrictions.

The remainder of this paper is organized as follows. Section 2 lays out and discusses
the econometric framework. Section 3 contains a simulation study. Section 4 applies the model to study the effects of a monetary policy shock. Section 5 concludes.

\section{ A large structural VAR with Non-Gaussian Factors}

Let $\boldsymbol{y}_t =(y_{1,t},\dots,y_{n,t})'$  be an $n \times 1$ vector of endogenous variables at time $t$. We write the model as
\begin{align*}
 \boldsymbol{y}_t&=\bm b_0 +\bm B_1 \bm y_{t-1}+\dots+\bm B_p \bm y_{t-p}+\bm u_t,\\
\bm u_t&=\boldsymbol{L}  \boldsymbol{f}_t+\boldsymbol{v}_t,
\label{VAR}
\end{align*}
where $\boldsymbol{v}_t \sim N(\boldsymbol{0},\boldsymbol{\Sigma})$ with $\boldsymbol{\Sigma}=\text{diag}(\sigma_1^2,\dots,\sigma_n^2)$, $\boldsymbol{f}_t$ is a $r\times 1$ vector, $\boldsymbol{L}$ is a $n\times r$ matrix and $\bm f_t \sim (\bm 0,\bm D)$ where $\bm D$ is a diagonal matrix. In more compact form the model can be written as 


\begin{equation}
\bm y_t=(\bm I_n \otimes \bm x'_t)\bm \beta+ \boldsymbol{L}  \boldsymbol{f}_t+\boldsymbol{v}_t
\label{reducedform}
\end{equation}

where $\bm I_n$ is the identity matrix of dimension $n$, $ \otimes$ is the Kronecker product, $\bm \beta=\text{vec}([\bm b_0,\bm B_1,\dots,\bm B_p]')$ and $\bm x_t=(1,\bm y'_{t-1},\dots,\bm y'_{t-p})'$ is a $k\times1$ vector of intercept and lagged values with $k=1+np$. The noise $\bm v_t$ could represent measurement error or other idiosyncratic factors. Heuristically, the $r$ factors are the structural shocks as they can affect more than one variable. Hence, we assume that the dynamics of $n$ variables are driven by $r$ structural shocks and noise $v_t$. This allows researcher to add variables to their model without adding additional structural shocks as would be the case in a standard VAR model. For example if a researcher wishes to an additional measure for inflation or output she does not wish to add an additional structural shock.

We follow \cite{Korobilis2022} and consider a reduced rank SVAR representation of the model. We obtain this representation by left-multiplying the reduced-form VAR model in (\ref{reducedform}) with the generalized inverse of $\bm L$, as follows:
\begin{align*}
\bm A \bm y_t&= \bm B \bm x_t+   \boldsymbol{f}_t+\bm A \boldsymbol{v}_t\\
\bm f_t&\approx\bm A \bm y_t-\bm B \bm x_t,
\end{align*}
where $\bm A=(\bm L'\bm L)^{-1}\bm L' $ and $\bm B=(\bm A\bm b_0,\bm A\bm B_1,\dots,\bm A\bm B_p)$.
By assumption the noise $\bm v_t$ is uncorrelated, the Central Limit Theorem in \cite{bai2003} suggests that for each $t$ and for $n\rightarrow \infty $ we have $\bm A \boldsymbol{v}_t\rightarrow 0 $ making the term asymptotically negligible. This justifies interpreting $\bm v_t$ as a noise shock which carries no structural interpretation. In contrast, the factors $\bm f_t$ have the interpretation as a projection of the SVAR structural shocks into $\mathbb{R}^r$. Therefore the model can be used to draw structural inference by using standard tools such as impulse response functions, see \cite{Froni2019}, \cite{Korobilis2022} and \cite{chan2022large}.\footnote{Compare also with the discussion of structural inference in dynamic factors models of chapter 16 in \cite{kilian2017structural}.}

Assuming that $\bm f_t$ and $ \bm v_t$ are uncorrelated we have $\text{Var}(\bm u_t|\bm \Sigma, \bm D)=\bm L\bm D \bm L'+\bm \Sigma$. Furthermore, to ensure one can separately
identify the common and the idiosyncratic components, we adopt a sufficient condition in \cite{anderson1956statistical} that  $r\leq (n-1)/2$. Precisely, for two observationally equivalent models such that
$\bm L\bm D\bm L'+\bm \Sigma =\bm L^*\bm D^* \bm L^{*}{'} +\bm \Sigma ^*$ it holds that $\bm L\bm D \bm L'=\bm L^*\bm D^*{'} \bm L^* $ and $\bm \Sigma =\bm \Sigma ^*$.
However, without additional restrictions the matrix of factor loadings $\bm L$ is not identified, i.e. any orthogonal matrix 
$\bm Q \in \mathcal{O}$ of the orthogonal group $\mathcal{O}=\{\bm Q \in\mathbb{R}: \bm Q \bm Q'=\bm I_r\} $
yields an observationally equivalent model $ \tilde{\bm L}  \tilde{\bm f_t}=\bm L \bm Q \bm Q' \bm f_t$. In the following, we discuss how independent non-Gaussian factors an be used to uniquely pin down the impact effect of the structural shocks, and hence archive point identification.  


\subsection{Identification by higher moments}
In this section we exploit information provided by higher moments to identify the model. 
To exploit this information we strengthen the assumptions that the factors $\bm f_t$ are uncorrelated with each other and uncorrelated with the noise $\bm v_t$ by assuming that the factors are also independent with each other and independent of the noise $\bm v_t$. Moreover, we assume that $\bm f_t$ and $\bm v_t$ have zero mean and finite moments up to the fourth order. These assumptions let us derive moment restrictions. In addition, we assume that $\bm L$ has full column rank and we can separately identify the common and the idiosyncratic components.\footnote{In the previous section we discuss that we need  $r\leq (n-1)/2$ to separate the noise from the common factors. If the factors are non-Gaussian and independent we can relax this condition. In particular,  \cite{bonhomme2009consistent} show that  if all factors are either skewed or kurtotic $r=n-1$ shocks can be identified and if all factors are kurtotic $r=n$ can be identified (in addition to some technical assumptions).} Multivariate cumulants of centred random variables of orders 2, 3 and 4 are defined as follows:
\begin{align*}
&\text{Cum}(Z_1,Z_2)=\mathbb{E}(Z_1,Z_2),\\
&\text{Cum}(Z_1,Z_2,Z_3)=\mathbb{E}(Z_1,Z_2,Z_3),\\
&\text{Cum}(Z_1,Z_2,Z_3,Z_4)=\mathbb{E}(Z_1,Z_2,Z_3,Z_4)-\mathbb{E}(Z_1,Z_2)\mathbb{E}(Z_3,Z_4)\\
&\quad \quad \quad \quad \quad \quad\quad-\mathbb{E}(Z_1,Z_3)\mathbb{E}(Z_2,Z_4)-\mathbb{E}(Z_1,Z_4)\mathbb{E}(Z_2,Z_3).
\end{align*}

Let $m \in \{2, 3,4\}$ and $ (i_1,\dots,i_m) \in (1,\dots,n)^m.$ Then we have

\begin{equation}
\text{Cum}(y_{i_1,t},\dots,y_{i_m,t})=\sum_{\tilde{r}=1}^r\left(\prod_{\tilde{m}=1}^m l_{i_{\tilde{m}},\tilde{r}} \right)\kappa_m(f_{k,t})+\text{Cum}(v_{i_1,t},\dots,v_{i_m,t}),
\label{restrictions}
\end{equation}
where we write $\kappa_m(Z)=\text{Cum}(Z,\dots,Z)$ (repeat $Z$ $m$ times) for univariate cumulants of order $m\geq1$. These moment restrictions have a common multilinear structure which allows us to write them in matrix form. Let us define the following $n\times n $, symmetric, square matrices:
\begin{align*}
\bm \Sigma_{y}&=[\text{Cum}(y_{i,t},y_{j,t})],\\
\bm \Gamma_y(\ell)&=[\text{Cum}(y_{i,t},y_{j,t},y_{\ell,t})], \quad \ell \in \{1,\dots,n\},\\
\bm \Omega_y(\ell,m)&=[\text{Cum}((y_{i,t},y_{j,t},y_{\ell,t},y_{m,t})], \quad \ell,m \in \{1,\dots,n\},\\
\end{align*}
with similar expressions for $\bm \Sigma_{v}$, $\bm \Gamma_v(\ell)$ and $\bm \Omega_v(\ell,m)$. Because 
$\boldsymbol{v}_t \sim N(\boldsymbol{0},\boldsymbol{\Sigma})$ we have that $\bm \Gamma_v(\ell)=\bm \Omega_v(\ell,m)=0$. Furthermore, we normalize by setting $\bm D=\bm I$. This choice is arbitrary as multiplication of the $k$th diagonal element just scales the $k$th column of $\bm L$. In practice, we normalize one element of the $k$th column of $\bm L$ (i.e. one impulse response to the $k$th shock in the impact period) to facilitate the economic interpretation, see section 4. Together with the restrictions in (\ref{restrictions}) this implies that
\begin{align}
\label {res1} \bm \Sigma_{y}&=\bm L\bm L'+\bm \Sigma,\\
\label {res2} \bm \Gamma_y(\ell)&=\bm L \bm K_3 \text{diag}(\bm L_{\ell})\bm L',\\
\label {res3} \bm \Omega_y(\ell,m)&=\bm L \bm K_4 \text{diag}(\bm L_{\ell}\odot \bm L_{m})\bm L',
\end{align}

where $\bm L_{\ell}$ is the $\ell$th row of $\bm L$, $\bm K_3$ (resp. $\bm K_4$) is the diagonal matrix with cumulant $\kappa_3(f_{k,t})$ (resp. $\kappa_4(f_{k,t})$) in the $k$th entry of the diagonal, and $\odot$ is the Hadamard (element by element) matrix product.

\begin{figure}[ht]
\centering 
\includegraphics[trim = 0mm 0mm 0mm 00mm,width=1\textwidth]{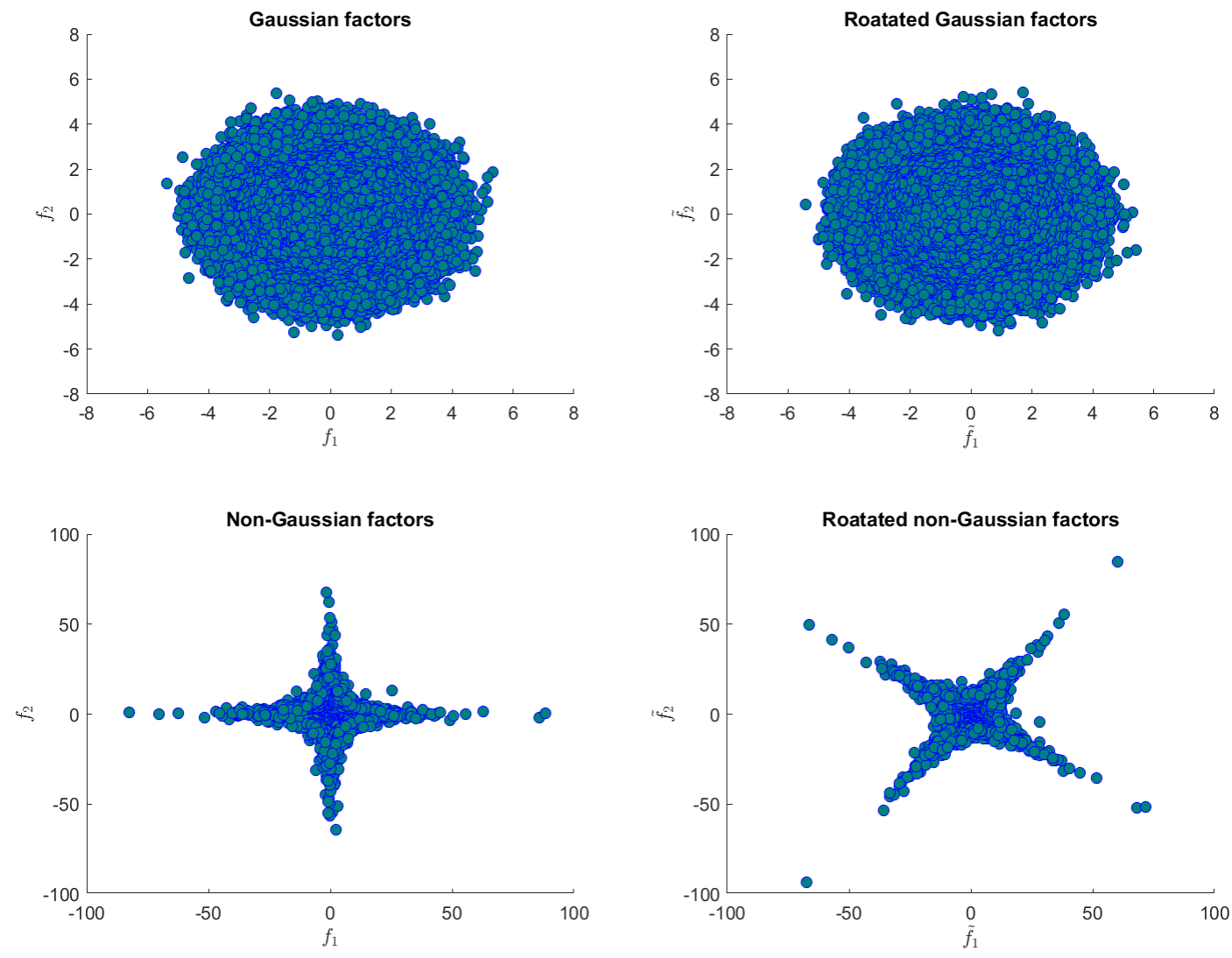} 
\caption[]{The figure illustrates how higher moments can provide information that can be exploited identification. In the upper part the factors are independently drawn from a normal distribution and in the lower part from a t-distribution with four degrees of freedom. The factors in the right part have been multiplied by an orthogonal matrix. }
\label{fig:Illustration} 
\end{figure}

Figure \ref{fig:Illustration} illustrate how higher moments provide information for the identification of factors and hence factor loadings. The figure plots the joint distribution of two factors $f_{1,t}$ and $f_{2,t}$. In the upper part they are independently drawn from univariate standard normal distributions and in the lower part they are drawn independently from univariate t-distributions with four degrees of freedom. In the right part of the figure, the factors have been multiplied with an orthogonal matrix as follows
\begin{gather}
 \begin{bmatrix} \tilde{f}_{1t}  \\ \tilde{f}_{2t} \end{bmatrix}
 =
  \begin{bmatrix}
   cos(\pi/5) &
     sin(\pi/5) \\
     -sin(\pi/5)  &
    cos(\pi/5) 
   \end{bmatrix}
	 \begin{bmatrix} f_{1t}  \\ f_{2t} \end{bmatrix}.
\end{gather}

Inspecting the upper part of figure \ref{fig:Illustration} reveals that the joint distribution of the Gaussian factors does not change. Indeed, the correlation between the factors and squared factors is zero before and after the rotation.  Inspecting the lower part of figure \ref{fig:Illustration} reveals that the joint distribution of the non-Gaussian factors changes after the rotation. Before the rotation the non-Gaussian factors are independent. After the rotation of the factors we can observe that a large value of one of the factors contains information about the other factors. In particular, while the factors are still uncorrelated, the squared factors are correlated after the rotation. Hence, the rotated factors are no longer independent. By utilizing the fact that the factors are independent, we can detect that the bottom right panel shows a rotation of the factors.

Formally, \cite{bonhomme2009consistent} proofs the following three points

\begin{enumerate}
\item If at most one factor variable has zero excess kurtosis, then factor loadings are identified from second- and fourth-order moments restrictions (\ref{res1}) and (\ref{res3}).
\item If at most one factor variable has zero skewness, then factor loadings are identified from second- and third-order moment restrictions (\ref{res1}) and (\ref{res2}).
\item If for any couple of factors indices $(k,k')$, $\kappa_3(f_{k,t}),\kappa_3(f_{k',t}),\kappa_4(f_{k,t}),\kappa_4(f_{k',t}))\neq 0$, then factor loadings are identified from second-, third- and fourth-order moment restrictions (\ref{res1}) to (\ref{res3}).
\end{enumerate}

\cite{bonhomme2009consistent} say that the factor loadings $\bm L$ are identified if the set of orthogonal matrices $\bm Q \in \mathcal{O}$ leading to observational equivalent models is reduced to the set of all products $\bm S \bm P$, where $\bm S$ is a diagonal matrix with diagonal components equal to $1$ or $-1$ and $\bm P$ is a permutation matrix. Thus,  $\bm L$ is identified up to sign switches of the columns and the order of the columns. Which of the different sign permutations we choose is arbitrary, as the economic interpretation of the results does not change. However, we need to be careful not to mix different different sign permutations when drawing from the posterior distribution. The Gibbs sampler algorithm we propose may sample from different sign permutations, such that posterior draws from the response of a variable to a shock do not come from a unique shock, but rather from a combination of different shocks, leading to invalid inference. However, the manifestation of the permutation problem in the posterior sample can be reliably diagnosed. For example, jumps between permutations should lead to multimodal posterior distributions, which are typically be easily observed by inspecting marginal posterior densities or trace plots, as argued by \cite{anttonen2024statistically}. Finally, the whole permutation problem is alleviated in the common case where there is only one shock of interest (e.g., a monetary policy shock), and for the analysis of any shock, only permutations with respect to the shock of interest need to be ruled out. Since factors and factor loadings are not sampled jointly in our Gibbs sampler, permutation switches are less likely to occur if the posterior distributions do not overlap. However, to rule out the possibility of permutation switches, we carefully inspect the posterior distributions, see section 4.2. In addition, we post-process the posterior draws. In particular, we calculate the correlation of the factors with the proxy of \cite{romer2004new} for each posterior draw and the factor. The factor with the highest correlation is ordered first and considered as the monetary policy shock, see \cite{bertsche2022identification} and \cite{lewis2021identifying}. Note that the reordering was not necessary in our empirical application. In the Monte Carlo study we address this issue by using the algorithm proposed by \cite{keweloh2024uncertain} to potentially reorder the columns of the factor loadings.

Importantly, the shocks in the model are not inherently structural and must be labeled manually by the researcher based on economic assumptions. These assumptions do not constrain the possible values of the identified parameters, which are derived purely from statistical information. Instead, the economic assumptions guide the selection among the statistically identified shocks, a desirable feature when the restrictions are considered approximate but not strictly valid (see \cite{lewis2024identification}). Section 4.2, describes in detail the process of labeling a monetary policy shock.

The assumption of independent structural shocks has been criticised by \cite{montiel2022svar}.
 \cite{montiel2022svar} argue that a potentially shared volatility process would violate this assumption. A shared volatility process would imply that multivariate cumulates of order 4 would no longer be all be zero and the moment restrictions in  (\ref{res3}) would be violated. In this case, we could replace the assumption of independence by assuming that multivariate cumulates of order 3 are all zero and still can use moment restrictions in  (\ref{res1}) and (\ref{res2}) to identify the factors loadings $\bm L$ without using the restrictions in (\ref{res3}). Of course, we could also replace the independence assumption by assuming that multivariate cumulates of order 4 are all zero.

%
%
\subsection{Prior Specifications}

We assume that the factors are independent and that $ f_{\tilde{r},t} \sim \mathcal{T}_{v_{\tilde{r}}}(0,1)$, where $\mathcal{T}_{v_{\tilde{r}}}(0,1)$ Student's distribution with zero mean, standard derivation of one and $v_{\tilde{r}}$ degrees of freedom for $\tilde{r}=1,\dots,r$. The degree of freedom parameter $v_{\tilde{r}}$ is treated as unknown and estimated from the data. We assume $v_{\tilde{r}} \sim \text{U}(2,30)$. Assuming $v_{\tilde{r}}>2$ ensures that the variance of the factors exists. The upper bound is set sufficiently high so that the $t$-distribution can, in principle, closely resemble the normal distribution. Therefore, as a special case, we allow that $\bm f_t \sim N(\bm 0,\bm I)$, which is often assumed. Thus, the data will inform us about deviations from Gaussianity. It is worth noting that although we use a symmetric prior distribution for the factors, our prior has fat tails and is updated by the likelihood function. Hence, the posterior distributions of the factors can be highly skewed if empirical warranted. This is what we observe in our empirical application.\footnote{We have also considered a skewed -t distribution as in \cite{karlsson2023vector} and find that this has little impact on the results.}

To facilitate computation we utilize a mixture representation of the t-distribution. Suppose $(X|\lambda) \sim N(\mu,\lambda \sigma^2)$, where $\lambda$ is a latent variable that scale the variance of $X$. Assume that $\lambda$ has an inverse-gamma distribution, particularly, $\lambda \sim IG(v/2,v/2)$, then the marginal distribution of $X$ is $\mathcal{T}_v(\mu,\sigma^2)$. Hence,  $\bm f_t \sim N(\bm 0,\bm W_t)$, with $\bm W_t=\text{diag}( w_{1,t},\dots,\ w_{r,t})$ and $w_{\tilde{r},t}\sim IG(v_{\tilde{r}}/2,v_{\tilde{r}}/2)$.

In high-dimensional settings such as large VARs, it is important to use shrinkage priors
to avoid overfitting. Next we describe the Minnesota-type adaptive hierarchical prior suggested by \cite{chan2024large}. This prior combines advantages of the Minnesota priors (e.g., rich prior beliefs such as cross-variable shrinkage) and modern adaptive hierarchical priors (e.g., heavy tails and substantial mass around the prior mean), see \cite{chan2021minnesota}.
Let $\bm \beta_i$ the VAR coefficients in the $i-$th equation, $i=1,\dots,n$. For $\beta_{i,j}$, the $j-$th coefficient in the $i-$th equation, let $\lambda_{i,j}=\lambda_1$ if it is a coefficient on an 'own lag' and let $\lambda_{i,j}=\lambda_2$ if it is a coefficient on an 'other lag'. Consider the prior for $\beta_{i,j}$, $i=1,\dots,n$ and $j=2,\dots,k$:

\begin{align}
\beta_{i,j}|\lambda_1, \lambda_2, \psi_{i,j} &\sim N(m_{i,j},\lambda_{i,j}\psi_{i,j}C_{i,j}),\\
\sqrt{\psi_{i,j}}&\sim  C^+(0,1),\\
 \sqrt{\lambda_1}, \sqrt{\lambda_2}&\sim C^+(0,1),
\end{align}
where $C^+(0,1)$ denotes the standard half-Cauchy distribution. The two hyperparameter $\lambda_1$ and $\lambda_2$ are the global variance components that are common to, respectively, coefficients of own and other lags, whereas each $\psi_{i,j}$ is a local variance component specific to the coefficients $\beta_{i,j}$. Furthermore, the prior mean $m_{i,j}$ is set to zero except for the coefficients associated with the first own lag, which is set to one. Lastly, the constants $C_{i,j}$ are obtained as in the Minnesota prior, i.e., $C_{i,j}=\frac{1}{p^2}$. If all local variances are fixed, i.e., $\psi_{i,j}=1$ the prior reduces to a Minnesota-typ prior. Therefore, the prior is an extension of the Minnesota prior by introducing local variance components such that the marginal prior distribution for $\beta_{i,j}$ has heavy tails to mitigate biases of large coefficients. On the other hand, if $m_{i,j}=0$, $C_{i,j}=1$ and $\lambda_1=\lambda_2$, then the prior reduces to the standard horseshoe prior where the coefficients have identical distributions, see \cite{carvalho2010horseshoe}. From this perspective, the prior can be viewed as an extension of the horseshoe prior which incorporates richer prior beliefs on the VAR coefficients, such as cross-variable shrinkage, i.e., shrinking coefficients on own lags differently than other lags, see \cite{chan2022asymmetric} for the empirical importance of cross-variable shrinkage.

To facilitate sampling, we follow \cite{makalic2015simple} and use the following latent variables representations of the half-Cauchy distributions:

\begin{align}
(\psi|z_{\psi_{i,j}})&\sim  \mathcal{IG}(1/2,1/z_{\psi_{i,j}}),\quad z_{\psi_{i,j}}\sim  \mathcal{IG}(1/2,1),\\
(\lambda_l|z_{\lambda_l})&\sim  \mathcal{IG}(1/2,1/z_{\lambda_l}),\quad z_{\lambda_l}\sim \mathcal{IG}(1/2,1),
\end{align}
for $i=1,\dots,n$, $j=2,\dots,k$ and $l=1,2$.

Finally, we present the prior distribution for the reaming model coefficients. Let $\bm l_i$ denote the elements of $\bm L$ in the $i-$th equation. We assume $\bm l_i \sim N(\bm l_{0,i},\bm V_{\bm l_{i}})$, and for the variance terms of the noise we assume $\sigma_{j}^2 \sim \mathcal{IG}(\alpha_0,\beta_0)$. We set $\bm l_{0,i}= \bm 0$, $ \bm V_{\bm l_i}=10  \times \bm I_r$ and $\alpha_0=\beta_0=0$.

\subsection{Gibbs Sampler}
In this section we develop an efficient posterior sampler to estimate the model. Posterior draws can be obtained by sampling sequentially from the conditional distributions:

\begin{enumerate}
\item $p(\bm f|\bm y,\bm \beta, \bm L,\bm \Sigma,\bm W,\bm v, \bm \lambda, \bm \psi, \bm z_{\lambda},\bm z_{\bm \psi})=p(\bm f|\bm y, \bm \beta, \bm L, \bm W, \bm \Sigma)$;
\item $p(\bm \beta, \bm L|\bm y, \bm f,\bm \Sigma,\bm W,\bm v, \bm \lambda, \bm \psi, \bm z_{\lambda},\bm z_{\bm \psi})=\prod_{i=1}^n=p(\bm \beta_i,\bm l_i|\bm y_i,\bm f, \bm \sigma^2_i)$
\item $p(\bm W|\bm y,\bm \beta, \bm L,\bm f,\bm \Sigma,\bm v, \bm \lambda, \bm \psi, \bm z_{\lambda},\bm z_{\bm \psi})=\prod_{\tilde{r}=1}^r\prod_{t=1}^T p(w_{\tilde{r},t}|v_{\tilde{r}},f_{\tilde{r},t})$;
\item $p(\bm v|\bm y,\bm \beta, \bm L,\bm f,\bm \Sigma,\bm W, \bm \lambda, \bm \psi, \bm z_{\lambda},\bm z_{\bm \psi})=\prod_{\tilde{r}=1}^r p(v_{\tilde{r}}|\bm W_{\tilde{r}})$;
\item $p(\bm \Sigma|\bm y,\bm \beta, \bm L,\bm f,\bm W,\bm v, \bm \lambda, \bm \psi, \bm z_{\lambda},\bm z_{\bm \psi})=\prod_i^n p(\sigma_i^2|\bm y_i,\bm f_i,\bm l_i,\bm \beta_i) $;
\item $p(\bm \lambda|\bm y,\bm \beta, \bm L,\bm f,\bm \Sigma,\bm W,\bm v, \bm \psi, \bm z_{\lambda},\bm z_{\bm \psi})=\prod_{l=1}^2p(\lambda_l|\bm \beta, \bm \psi, z_{\lambda_l})$;
\item $p(\bm \psi|\bm y,\bm \beta, \bm L,\bm f,\bm \Sigma,\bm W,\bm v, \bm \lambda, \bm z_{\lambda},\bm z_{\bm \psi})=\prod_{i=1}^2\prod_{j=2}^k p(\psi_{i,j}|\beta_{i,j},\bm \lambda,z_{psi_{i,j}})$;
\item $p( \bm z_{\lambda}|\bm y,\bm \beta, \bm L,\bm f,\bm \Sigma,\bm W,\bm v, \bm \lambda, \bm \psi,\bm z_{\bm \psi})=\prod_{l=1}^2p(z_{\lambda_l}|\lambda_l)$;
\item $p(\bm z_{\bm \psi}|\bm y,\bm \beta, \bm L,\bm f,\bm \Sigma,\bm W,\bm v, \bm \lambda, \bm \psi, \bm z_{\lambda})=\prod_{i=1}^2\prod_{j=2}^k p(z_{ \psi_{i,j}}| \psi_{i,j})$,
\end{enumerate}

with $\bm y_i=(y_{i,1},\dots,y_{i,T})'$ be a $T\times 1$vector of observations of the $i-$th variable and $\bm W_{\tilde{r}}=( w_{\tilde{r},1},\dots,\ w_{\tilde{r},T})$.

\textbf{Step 1} First, we sample $\bm f_t$. We stack $\bm y= ( \bm y_1',\dots ,\bm y_T')'$, $\bm f=(\bm f_1',\dots, \bm f_T')'$ and write the model in compact form as

\begin{equation}
\bm y=\bm X \bm \beta+ (\bm I_T \otimes \bm L)\bm f+\bm v,  \qquad \bm v \sim N(\bm 0,\tilde{\bm \Sigma} ),
\label{compactformF}
\end{equation}
where $\tilde{\bm \Sigma}=\bm I_T \otimes \bm \Sigma$ and $\bm X$ is the matrix of intercepts and lagged values. From the mixture representation it follows that $(\bm f|\bm W)\sim N(\bm 0, \bm W)$ with $ \bm W=\text{diag}(\bm W_1,\dots,\bm W_T)$. Then we can use standard regression results (see, e.g., \cite{chan2019bayesian}) to obtain

\begin{equation}
(\bm f|\bm y,\bm \beta, \bm L, \bm W)\sim N(\hat{\bm f},\bm K_{\bm f}^{-1}),
\end{equation}

where

\begin{equation}
\bm K_{\bm f}=\bm W^{-1}+ (\bm I_T \otimes \bm L')\tilde{\bm \Sigma}^{-1} (\bm I_T \otimes \bm L), \quad \hat{\bm f}=\bm K_{\bm f}^{-1}(\bm I_T \otimes \bm L')\tilde{\bm \Sigma}^{-1}(\bm y-\bm X \bm \beta).
\end{equation}
It is worth mentioning that $\bm K_{\bm f}$ is a band matrix and because of this one can use the precision sampler of \cite{chan2009efficient} to sample $\bm f$ efficiently.

\textbf{Step 2} Second, we sample $\bm \beta$ and $\bm L$ jointly to improve sampling efficiency. Given the latent factors $\bm f$, the VAR becomes $n$ unrelated regressions and we can sample $\bm \beta$ and $\bm L$ equation by equation. Equation by equation estimation simplifies the estimation and allows for the estimation with a large number of variables. Remember that $\bm \beta_i$ and $\bm l_i$ denote, respectively, the VAR coefficients and the factor loadings in the $i-$th equation. Then, the $i-$th equation of the VAR can be expressed as

\begin{equation}
\bm y_i=\bm X_i \bm \beta_i+\bm F \bm l_i +\bm v
\end{equation}
where $\bm F=(\bm f_1,\dots,\bm f_r)$ the $ T\times r$ matrix of factors with $ \bm f_i=(f_{i,1},\dots,f_{i,T})'$. The vector of noise $\bm v=(v_{i,1},\dots,v_{i,T})'$ is distributed as $N(\bm 0,\bm I_T \sigma_i^2)$. We can write it more compactly by defining $\bm \theta_i=(\bm \beta_i',\bm l_i')'$ and $\bm Z_i=(\bm X_i, \bm F)$,

\begin{equation}
\bm y_i=\bm Z_i \bm \theta_i+\bm F \bm l_i +\bm v.
\end{equation}

Then using standard linear regression results, we get
\begin{equation}
(\bm \theta_i|\bm y_i,\bm f,\sigma_i^2)\sim N( \hat{\bm \theta},\bm K^{-1}_{\bm \theta_i})
\end{equation}
where
\begin{align*}
\bm K_{\bm \theta_i}=\bm V^{-1}_{\bm \theta_i}+\bm \sigma^{-2}_i \bm Z_i'\bm Z_i, \quad  \hat{\bm \theta}_i=\bm K^{-1}_{\bm \theta_i}(\bm V^{-1}_{\bm \theta_i}\bm \theta_{0,i}+\sigma^{-2}_i \bm Z_i \bm y_i)
\end{align*}
with $\bm V_{\bm \theta_i}=\text{diag}(\bm V_{\bm \beta_i},\bm V_{\bm l_i})$ with $\bm V_{\bm \beta_i}=\text{diag}(C_{i,1},\lambda_{i,2}\psi_{i,2}C_{i,2},\dots,\lambda_{i,k}\psi_{i,k}C_{i,k})$ and $\bm \theta_{0,i}=(\bm m_i',\bm l_{0,i})'$ with $\bm m_i=(m_{i,1},\dots,m_{i,k})'$.

\textbf{Step 3} We sample the latent variable $\bm W_{\tilde{r}}$. The posterior is proportional to 

\begin{equation}
p(\bm W_{\tilde{r}}|\boldsymbol{f}_{\tilde{r}},v_{\tilde{r}}) \propto \prod_{t=1}^{T} \left[(\bm w_{\tilde{r},t})^{-(\frac{v_{\tilde{r}}+1}{2}+1)} \text{e}^{-\frac{1}{2w_{\tilde{r},t}}\left(v_{\tilde{r}}+f_{\tilde{r},t}^2\right)}  \right],
\end{equation}
which is a product of inverse-gamma kernels. Therefore, we conclude

\begin{equation}  
w_{\tilde{r},t}|f_{\tilde{r},t},v_{\tilde{r}}, \sim \mathcal{IG}\left(\frac{v_{\tilde{r}}+1}{2},\frac{1}{2}\left(v_{\tilde{r}}+f_{\tilde{r},t}^2\right) \right).
\end{equation}

\textbf{Step 4} To sample from $v_{\tilde{r}}|\bm W \propto \mathcal{U}(2,30)\prod_{t=1}^T \mathcal{IG}(w_{\tilde{r},t};\frac{v_{\tilde{r}}}{2},\frac{v_{\tilde{r}}}{2})$ we use a Griddy-Gibbs sampler as this conditional density of $v_{\tilde{r}}$ is nonstandard. The idea is to use a inverse-transform method.
As the inverse of the target density is not available analytically we 
use Griddy-Gibbs sampler to approximating sampling from univariate distributions with bounded support.
It is basically a discretized version of the inverse-transform method and only requires the evaluation of the density (up to a normalizing constant).
We construct an approximation of the density function of $v_{\tilde{r}}$ on a fine grid. 
Given the discretized density, we can implement the inverse-transform method for a discrete random variable. We wish to sample $v$ with density $f$ and bounded support on $(a,b)$. In our case $a=2$ and $b=30$. The Griddy Gibbs algorithm proceeds as follows:
\begin{enumerate}
\item Construct a grid with grid points $v_1,\dots,v_n$, where $v_1=a$ and $v_n=b$.
\item Compute $F_i=\sum_{j=1}^i f(v_j)$.
\item Generate $U$ from $\mathcal{U}(0,1)$.
\item Find the smallest positive integer $q$ such that $F_q\geq U$ and return $v=v_q$.
\end{enumerate}

\textbf{Step 5} Next, we sample $\sigma^2_{i}$ for $i1,\dots,n$. Given $\bm f$ the model reduces to $n$ independent linear regressions. Therefore, we can use standard regression results (see, e.g., \cite{chan2019bayesian}) to obtain

\begin{equation}
(\sigma_{i}^2|\bm y, \bm f, \bm L)\sim \mathcal{IG}\left(\alpha_0+\frac{T}{2},\beta_0+0.5\sum_{t=1}^T(y_{it}-\bm X_{it}\bm\beta_i -\bm l_i\bm f_t)^2\right).  
\end{equation}

\textbf{Step 6} Lastly, we sample the hyperparameter $\lambda_{1}$, $\lambda_{2}$ and $\psi_{i,j}$ from our shrinkage prior for the VAR coefficients as well as the mixing variables $z_{\lambda_{1}}$, $z_{\lambda_{2}}$ and $z_{\psi_{i,j}}$. Using the latent variable representation of the half Cauchy distribution, we obtain

\begin{align*}
p(\psi_{i,j}|\beta_{i,j}, \lambda_{i,j}, z_{\psi_{i,j}})&\propto \psi_{i,j}^{\frac{1}{2}} \text{e}^{-\frac{1}{2\lambda_{i,j}C_{i,j}\psi_{i,j}}(\beta_{i,j}-m_{i,j})^2  }\times \psi^{-\frac{3}{2}}\text{e}^{-\frac{1}{\psi_{i,j}z_{\psi_{i,j}}}}\\
&=\psi_{i,j}^{-2}\text{e}^{-\frac{1}{\psi_{i,j}}\left(\frac{1}{z_{\psi_{i,j}}}+\frac{(\beta_{i,j}-m_{i,j})^2}{2\lambda_{i,j}C_{i,j}}\right)},
\end{align*}
which is the kernel of the following inverse-gamma distribution:
\begin{equation}
(\psi_{i,j}|\beta_{i,j}, \lambda_{i,j}, z_{\psi_{i,j}})\sim \mathcal{IG}\left(1, \frac{1}{z_{\psi_{i,j}}}+\frac{(\beta_{i,j}-m_{i,j})^2}{2\lambda_{i,j}C_{i,j}}\right).
\end{equation}

We denote $S_{\lambda_1}$ as a collection of all the indexes $(i,j)$ such that $\beta_{i,j}$ is a coefficient associated with an own lag. More precisely, $S_{\lambda_1}=\{(i,j): \beta_{i,j} \quad \text{is a coefficient associated with an own lag}\}$. Similarly, define $S_{\lambda_2}$ as the set that contains all the indexes $(i,j)$ such that $\beta_{i,j}$ is a coefficient associated with a lag of other variables. Then, we have

\begin{align*}
p(\lambda_1|\bm \beta, \bm \psi, z_{\lambda_1})&\propto \prod_{(i,j)\in S_{\lambda_1}} \lambda_1^{-\frac{1}{2}} \text{e}^{-\frac{1}{2\lambda_1 C_{i,j}\psi_{i,j}}(\beta_{i,j}-m_{i,j})^2   }\times \lambda_1^{-\frac{3}{2}} \text{e}^{-\frac{1}{\lambda_1 z_{\lambda_1}}  },\\
&=\lambda_1^{-\left(\frac{np+1}{2}+1 \right)}
\text{e}^{-\frac{1}{\lambda_1}\left(\frac{1}{z_{\lambda_1}} +\sum_{(i,j)\in S_{\lambda_1}} \frac{(\beta_{i,j}-m_{i,j})^2}{2\psi_{i,j}C_{i,j}}  \right)   },
\end{align*}

 which is the kernel of the following inverse-gamma distribution:
\begin{equation}
(\lambda_1|\bm \beta, \bm \psi, z_{\psi_{\lambda_1}})\sim \mathcal{IG}\left(\frac{np+1}{2}, \frac{1}{z_{\lambda_1} }+\sum_{(i,j)\in S_{\lambda_1}}\frac{(\beta_{i,j}-m_{i,j})^2}{2\psi_{i,j}C_{i,j}}\right).
\end{equation}
Similarly, we have
\begin{equation}
(\lambda_2|\bm \beta, \bm \psi, z_{\psi_{\lambda_2}})\sim \mathcal{IG}\left(\frac{np+1}{2}, \frac{1}{z_{\lambda_2}}+\sum_{(i,j)\in S_{\lambda_2}}\frac{(\beta_{i,j}-m_{i,j})^2}{2\psi_{i,j}C_{i,j}}\right).
\end{equation}

 Furthermore, we sample the latent variables $\bm z_{\bm \psi}$ and $\bm z_{\bm \lambda}$. In particular, $z_{\psi_{i,j}}\sim \mathcal{IG}(1,1+\psi_{i,j}^{-1})$ for $i=1,\dots,n$ and $j=2,\dots,n$. Similarly, we have $z_{\lambda_{l}}\sim  \mathcal{IG}(1,1+\lambda_l^{-1})$ for $l=1,2$.

\subsection{DIC Estimation}
In complex hierarchical models such as ours, basic concepts such as parameters and their dimensions are
not clearly defined. In their seminal
paper, \cite{spiegelhalter2002bayesian} introduce the concept of effective number of parameters
and develop the theory of the DIC criteria for model comparison.\footnote{\cite{Korobilis2022} argues that for the purpose of assessing the fit
of a VAR that is intended to be used for impulse responses the DIC can be considered as more appropriate compared to alternative in-sample measures of fit.}
The model selection criterion is based on the deviance, which is defined as

\begin{equation}
D(\bm \theta)=-2\text{log}f(\bm y|\bm \theta)+2 \text{log} h(\bm y)
\end{equation}

where $f(\bm y|\bm \theta)$ is the likelihood function of the parametric model with parameter vector $\bm \theta)$ and $h(\bm y)$ is a function from the data alone and for model comparison set to $h(\bm y)=1$. The effective number of parameters $p_D$ is defined as

\begin{equation}
p_D=\overline{D(\bm \theta)}-D(\tilde{\bm \theta}),
\end{equation}

where $ \overline{D(\bm \theta)}= 2 \mathbb{E}_{\theta}(\text{log} f(\bm y|\bm \theta)$ is the posterior mean deviance and $\tilde{\bm \theta}$ is an estimate of $\bm \theta$, which is typically taken as the posterior mean or median. Heuristically, the effective number of parameters measures the
reduction in uncertainty due to estimation. The larger the reduction,
the more complex the model is. Then, the deviance information criterion is defined as

\begin{equation}
\text{DIC}=\overline{D(\bm \theta)}+2p_D.
\end{equation}
 
Given a set of models , the preferred model is the one with the minimum DIC value. It is clear from the above definition that the DIC depends on the prior only via its
effect on the posterior distribution. In situations where the likelihood information
dominates, one would expect that the DIC is insensitive to different prior distributions.

\cite{Celeux2006} point out that there are different alternative definitions of the DIC depending on different concepts of the likelihood. For example, let $\bm z$ denote a vector of latent variables then the integrated likelihood $f(\bm y|\bm \theta)$ is related to the conditional likelihood $f(\bm y|\bm z, \bm \theta)$ via

\begin{equation}
p(\bm y|\bm \theta)=\int p(\bm y|\bm \theta,\bm z) p(\bm z|\bm \theta)\text{d}\bm z.
\end{equation}

The DIC can then be defined based on the conditional likelihood instead of the integrated likelihood. The advantage of the DIC based on the conditional likelihood is that it is available in closed form for our model and is easy to evaluate. However, some papers have warned against using conditional likelihood version as a model comparison criterion for both theoretical and practical reasons. \cite{li2013robust} argue that the conditional likelihood of the augmented data
is non-regular and thus invalidates the standard asymptotic arguments used to justify the original DIC. On practical grounds, \cite{Miller2009} and \cite{chan2016observed} provide Monte Carlo evidence that this variant of the DIC almost always
favours the most complex models. Therefore, we next integrate out the latent variables from our model to evaluate the integrated likelihood and to compute the DIC. Conditioning on the mixing variables $\bm W_t$, the factors $\bm f_t$ and the noise $\bm v_t$ are jointly Gaussian

\[\begin{pmatrix}
\bm v_t \\
\bm f_t
\end{pmatrix}\sim \mathcal{N} \left(\begin{pmatrix}
\bm 0 \\
\bm 0
\end{pmatrix},\begin{pmatrix}
\bm \Sigma & \bm 0 \\
\bm 0 & \bm W_t
\end{pmatrix}\right).
\]

Then the conditional distribution of $\bm y$ given $\bm W$ but marginal of $\bm f$ has the analytic expression
\begin{align*}
p(\bm y|\bm \beta, \bm L, \bm W, \bm \Sigma)&=(2 \pi)^{-\frac{Tn}{2}} \prod_{t=1}^T|\bm L \bm W_t\bm L'|^{\frac{1}{2}} \text{e}^{-\frac{1}{2}(\bm y_t-(\bm I_n \otimes \bm x_t')\bm \beta)' 
(\bm L \bm W_t\bm L')^{-1}(\bm y_t-(\bm I_n \otimes \bm x_t')\bm \beta) }
\end{align*}

Next, the integrated likelihood can be written as

\begin{align*}
p(\bm y| \bm \beta, \bm L, \bm \Sigma, \bm v)=\int \frac{p(\bm y|\bm \beta,\bm L,\bm W,\bm \Sigma)p(\bm W|\bm v)}{g(\bm W)}g(\bm W) \text{d}\bm W.
\end{align*}

Hence, we can evaluate the integrated likelihood via importance sampling:

\begin{equation}
\hat{p}(\bm y| \bm \beta, \bm L, \bm \Sigma, \bm v)=\frac{1}{R}\sum_{r=1}^{R}\frac{p(\bm y|\bm \beta,\bm L,\bm W^{(r)},\bm \Sigma)p(\bm W^{(r)}|\bm v)}{g(\bm W^{(r)})},
\label{importancesampler}
\end{equation}
where $\bm W^{(1)},\dots, \bm W^{(R)}$ are draws from the importance distribution $g$. The quality of the importance sampling density estimator in (\ref{importancesampler}) depends on the choice of the the importance distribution. The conditional density of the latent variables $p(\bm W|\bm y, \bm \beta, \bm L, \bm \Sigma,\bm v)\propto p(\bm y|\bm \beta,\bm L,\bm W,\bm \Sigma)p(\bm W|\bm v)$ leads to a zero-variance importance estimator. While this density is unknown it provides guidance for choosing a good importance density. In particular, we wish to select $g(\bm W)$ "`close"' to the optimal density $f^* \propto p(\bm W|\bm y, \bm \beta, \bm L, \bm \Sigma,\bm v)$. We follow \cite{ChanEisenstat2015} to use the improved cross-entropy method to construct the importance density.

Consider a parametric family $\mathcal{F}=\{f(\bm W; \bm \upsilon)\}$ indexed by a parameter vector $\bm \upsilon$ within which we locate the importance density which is "`closest"' to the optimal importance density. The Kullback-Leibler divergence (or called cross entropy) is one convenient measure of closeness between densities. In particular, let $h_1$ and $h_2$ be two probability density functions. Then, the Kullback-Leibler distance from $h_1$ to $h_2$ is defined as

\begin{equation}
\mathcal{D}(h_1,h_2)=\int h_1(\bm x)\text{log} \frac{h_1(\bm x)}{h_2(\bm x}d \bm x.
\end{equation} 

Given this measure of closeness, we select the density $f(\cdot; \bm \upsilon) \in \mathcal{F}$ such that $\mathcal{D}(f^*,f(\cdot; \bm \upsilon))$ is minimized, i.e. $\bm \upsilon^*=\text{argmin}_{\bm \upsilon}\mathcal{D}(f^*,f(\cdot; \bm \upsilon))$. The solution of this minimization problem can be shown to be equivalent to finding

\begin{equation}
\bm \upsilon^*=\text{argmin}_{\bm \upsilon}\int p(\bm W|\bm y, \bm \beta, \bm L, \bm \Sigma,\bm v)\text{log} f(\bm W; \bm \upsilon)d\bm W
\end{equation}

This optimization problem is difficult to solve analytically, Instead, we consider the stochastic counterpart:

\begin{equation}
\hat{\bm \upsilon}^*=\text{argmin}_{\bm \upsilon} \frac{1}{M}\sum_{m=1}^M \text{log} f(\bm W_m;\bm \upsilon),
\label{MiniDevproblem}
\end{equation}
where $\bm W_1, \dots,\bm W_M$ are draws from the density $p(\bm W|\bm y, \bm \beta, \bm L, \bm \Sigma,\bm v)$. Hence, $\hat{\bm \upsilon}^*$ is the maximum likelihood estimate for $\bm \upsilon$ if we use $f(\bm W_m;\bm \upsilon)$ as the likelihood function with parameter vector $\bm \upsilon$ and $\bm W_1, \dots,\bm W_M$ as our observed data.  We consider the parametric family

\begin{equation}
\mathcal{F}=\Biggl\{ \prod_{t=1}^T \prod_{\tilde{r}=1}^r f_{\mathcal{IG}}(w_{\tilde{r},t},\alpha_{\tilde{r},t},\beta_{\tilde{r},t})\Biggr\},
\end{equation}

where $f_{\mathcal{IG}}$ is a inverse Gamma density. Given this choice of parametric family, the minimization problem in (\ref{MiniDevproblem}) can be solved using standard routines. In addition, we can use the Gibbs sampler of the joint posterior to obtain draws of  $\bm W_1, \dots,\bm W_M$ as we only need to be able to obtain draws from the marginal distribution given this choice of parametric family.

\subsection{Adding Economic Restrictions}

In this section we discuss how we can add economic restrictions such as zero restrictions, sign restrictions and proxy variables to our model framework.  Since our model is identified by higher moments, these restrictions are over-identifying restrictions. Combining identification based on higher moments with identification motivated by economic knowledge offers a number of attractive features. We can incorporate economic information to strengthen identification by higher moments, see \cite{carriero2024blended}. \cite{montiel2022svar} argue that inference based on higher moments necessarily demands more from a finite sample than identification based on economically motivated restrictions. Short-run restrictions, sign restrictions or instrumental variables can help when the conditions for point identification through statistical identification are not met and can help when higher moments provide only weak identifying information to improve estimation properties, see \cite{keweloh2023estimating}. In addition, identification based on economic prior knowledge provides natural solutions to the labelling problem, \cite{braun2023importance}.
Moreover, identification based on higher moments can be used to check economically motivated restrictions against the data. We can do this by using posterior summaries of the model parameters directly, or by comparing the DIC of the unrestricted model with the model in which we impose the economic restrictions.

Zero restrictions can be added on $\bm l_i$ by redefining $\bm l_i$ and $\bm F$ appropriately. For example, if the first element of $l_i$ is restricted to be zero, we can define $\tilde{l}_i$ to be the vector consisting of the second to $r$-th elements of $\bm l_i$ and $\tilde{ \bm F}=(\bm f_2,\dots, \bm f_r)$. Then, we replace $\bm F \bm l_i$ by $\tilde{\bm F}\tilde{\bm l}_i$. Sign restrictions can be implemented by drawing $L$ from a truncated multivariate normal distribution using the algorithm proposed by \cite{botev2017normal}. The algorithm does not scale well to higher dimension and we may want draw $\bm L$ conditional on $\bm \beta$ to speed up computation, see \cite{Korobilis2022} and \cite{chan2022large}. Finally, a proxy variable $m_t$ can be incorporated by adding one equation to the model in (\ref{reducedform}):
\begin{equation}
m_t= \tilde{\bm L} \bm f+ \tilde{v}_t,
\end{equation} 

see \cite{banbura2023drives}. An instrument is said to be valid if it is correlated with the shock of interest which we aim to identify and uncorrelated with all other shocks, see \cite{mertens2013dynamic}. We can impose the second assumptions by placing zero restrictions on  $\tilde{\bm L}$, see \cite{caldara2019monetary}. Given that the first factor  $f_{1,t}$ is the shock of interest we have that $ \tilde{\bm L}=(\tilde{l}_1, 0_2,\dots,0_r)$.

\section{Experiments with Artificial Data}
In this section, we evaluate the frequentist estimation
properties of the non-Gaussian factor model in a Monte Carlo study. The data generating process is $\bm y_t=\bm L \bm f_t +\bm v_t$ where

\begin{equation}
\bm L'= \left( \begin{array}{rrrrr rrrrr rrrr}
0 & 1 & 1& 1 & 1   & -1 & -1 & 1 & 1&1 & 1 & 1& 1&1\\
1 & 1 & 1& -1 & -1   & 1 & -1 & -1 & -1&1 & -1 & 1& 1&1\\
-1 & -1 & -1& -1 & -1   & 1 & -1 & -1 & -1&1 & -1 & 1& -1&-1\\
\end{array}\right),
\label{DGPL}
\end{equation}
$v_t\sim N(\bm 0, \bm I)$ and the factors $\bm f_t$  are drawn independently and identically either from a t-distribution with mean zero, variance one and four degree of freedom or from a pearson distribution
with mean zero, variance one, skewness 0.68 and excess kurtosis 15. We generate 1000 data sets with $T=500$ and $T=1000$ observations.

 Table \ref{MC} shows the bias, the mean squared estimation error (MSE), the average length of 68\% credible bands and the coverage rate (defined as the proportion of credible bands containing the true value). To save space we show the results for the first four elements of the first column of equation \ref{DGPL}). For both distributions the model is able to provide unbiased estimates and the correct coverage rate (the coverage rate is close to the probability
chosen for the credible bands). Furthermore, the estimation accuracy and the estimation precision are reasonable and increase with increasing sample size for both distributions. This shows that the model has good estimation properties for different distributions of the factors. As our prior follows a t-distribution, the estimation accuracy in terms of MSE is better and the estimation precision in terms of smaller credible bands are better if the shocks are generated by the t-distribution compared to the person distribution. However, it is plausible that these differences become smaller as the sample size increases.

\begin{table}[t]
\center
\caption{Simulation Results}
\begin{tabular}{lrrrr c rrrr} 
\toprule
&\multicolumn{4}{ c }{T=500}&&\multicolumn{4}{ c }{T=1000} \\
\cline{2-5} \cline{7-10}
   &Bias$$&MSE&Length&Coverage&&Bias&MSE&Length&Coverage\\ \toprule
\multicolumn{3}{l }{t-distribution}&& \\ \midrule
$l_{1,1}$& $-0.0014$ & $0.0091$ & $0.2026$&0.7050  && $0.0005$ & $  0.0043 $&$0.1291 $& 0.6700 \\
$l_{2,1}$& $0.0097$ & $0.0213$ & $0.2857$&0.7160  && $0.0026$ & $ 0.0084$&$0.1824$&0.6840\\
$l_{3,1}$& $0.0059$ & $ 0.0208$ & $0.2852$&0.6990  && $0.0013$ & $0.0082$&$0.1822 $&0.6950\\
$l_{4,1}$& $0.0024$ & $0.0025$ &0.0978& 0.7010  && $-0.0008$ & $ 0.0012 $&$0.0667$&0.6740\\ \toprule
\multicolumn{3}{l }{Pearson distribution}&& \\ \midrule
$l_{1,1}$& $-0.0030$ & $0.0176 $ & $0.2806  $&0.6960  && $-0.0026 $ & $0.0065$&$ 0.1655$&0.6830    \\
$l_{2,1}$& $ 0.0229$ & $ 0.0383  $ & $0.3997$&0.6940  && $  0.0042$ & $0.0130$&$ 0.2327$& 0.7050\\
$l_{3,1}$& $0.0239$ & $0.0397$ & $0.4004$& 0.6800 && $  0.0028$ & $ 0.0127$&$  0.2326$&  0.6990\\
$l_{4,1}$& $-0.0004$ & $0.0051$ &0.1478&  0.6780 && $ 0.0006$ & $0.0025$&$0.0987$& 0.6760 \\ \toprule
\end{tabular} 
\label{MC}
\begin{tabular}{p{15cm}}
\footnotesize  \textit{Notes}: The table shows the Bias, mean squared estimation error (MSE), average length of 68\% credible bands and the coverage rate (defined as the proportion in which the credible bands contain the true value). The factors are drawn independently and identically either from a t-distribution or person distribution. 
\end{tabular}
 \end{table}

\section{Empirical Application to Monetary Policy}
In this section we apply our model to identify a monetary policy shock using information from higher moments. Overall, the empirical analysis highlights the benefits of including more variables and performing a more comprehensive structural analysis. We begin with a discussion of the data and the model specification. The shocks are statistical identified, but the researcher needs to attach an economic meaning to them. We therefore present different ways of labelling the monetary policy shock, all of which lead to the same conclusion. We then assess the empirical plausibility of assuming non-Gaussian and mutually independent structural shocks. The analysis of impulse response functions shows that both prices and output respond with a large delay to a monetary policy shock. Finally, we illustrate how we can add a proxy variable to the model and use the DIC to check the empirical validity of exogenous exclusion restrictions.

\subsection{Data and Model Specification} 

We use the six variables used by \cite{uhlig2005effects}.  \cite{uhlig2005effects} uses real gross domestic product, the GDP deflator, a commodity price index, the total reserves, non-borrowed reserves and the federal funds rate. We extend this dataset to include nine additional variables. These include various measures of prices, economic activity, and variables representing the financial and labour markets.\footnote{We end up with 15 endogenous variables, the same number of variables as used in \cite{Korobilis2022}. Although we could certainly add even more variables, we consider the model to be reasonably large. } The data range from 1969M1 to 2007M12. Table \ref{tab:Data} contains detailed information on the variables, their sources, abbreviations and transformations. All variables are standardised for the estimation. We also use the exogenous measure of the US monetary policy shock from \cite{romer2004new} as a proxy variable. \cite{romer2004new} usese detailed
quantitative and narrative records to infer the Federal Reserve's intentions concerning the federal funds rate around FOMC meetings to develop an exogenous measure of the US monetary policy shock for our sample period.  Although \cite{romer2004new} themselves state
that their series is only "relatively free of endogenous and anticipatory movements" it is reasonable to use it to label the monetary policy shock. In line with the monthly frequency of the data, we follow  \cite{uhlig2005effects} and estimate the model with $p=12$ lags. The number of shocks $r$ is chosen according to the DIC. Table \ref{tab_DIC} shows the DIC for different numbers of shocks. The DIC favours the model with $r=4$. However, our empirical results are very robust to decreasing or increasing the number of shocks.

\begin{table}[t]
	\centering
		\captionof{table}{Deviance Information Criteria for the number of factors}
		     \begin{threeparttable}
	\begin{tabular}{cccc}
		\toprule
	\textbf{$r=3$} &\textbf{$r=4$} &\textbf{ $r=5$}  &\textbf{$r=6$} \\
		\midrule
       -119986 & -167567 &-157563 & -155730 \\
		\bottomrule
	\end{tabular}
 \begin{tablenotes}\footnotesize
    \item[]
    Notes: The table contains the DIC for different number of factors. Small values are preferred.
    \end{tablenotes}
		 \end{threeparttable}
	\label{tab_DIC}
\end{table}
\vspace{-10pt}

\subsection{Labelling the Monetary Policy Shock}

\begin{figure}[t]
	\centering
	\includegraphics[width=1\linewidth]{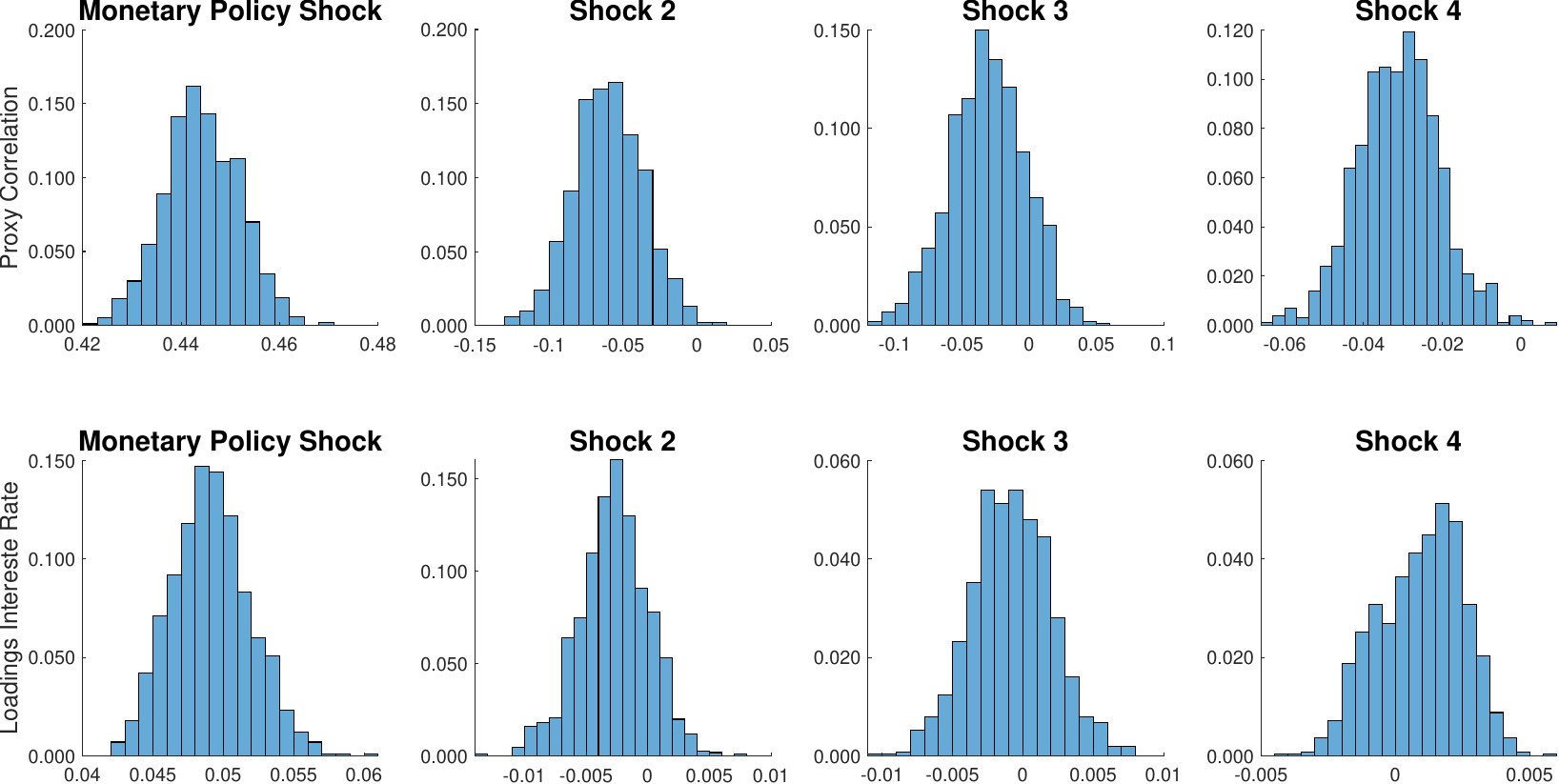}
	\caption{The fist panel shows the posterior distribution of the correlation of each shock with the proxy variable of \cite{romer2004new}. The second panel shows the posterior distribution of the loadings in the interest rate equation of each shock.  }
	\label{fig:proxyandLoadings}
\end{figure}

\begin{figure}[t]
	\centering
	\includegraphics[width=1\linewidth]{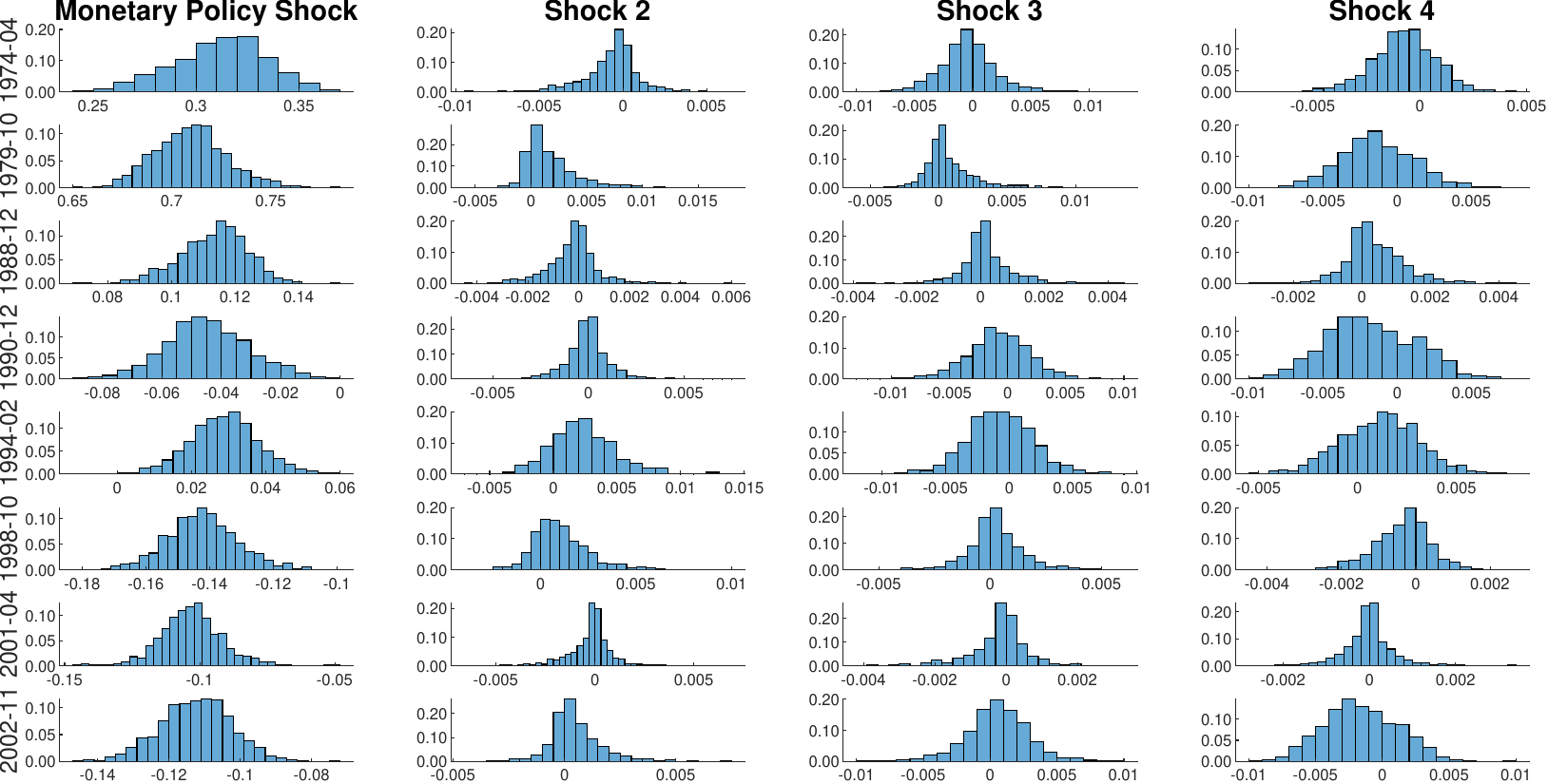}
	\caption{The figure shows the posterior distributions of the product of posterior loadings of the interest rate equation times each shock at specific points in time. }
	\label{fig:Narratives}
\end{figure}

Identification by higher moments leads to identification from a statistical point of view. But the research needs to attach an economic meaning to these shocks. Next, we discuss how we label a monetary policy shock using economic reasoning. First, \cite{lanne2023identifying} argues that a monetary policy shock should lead to an interest rate hike on impact. The lower part of figure \ref{fig:proxyandLoadings} plots the posterior distributions of the loadings of the interest rate equation. Only one of the shocks has a clear positive impact on the interest rate. Therefore, from an economic perspective the other shocks are no candidates for a monetary policy shock. Second, a monetary policy shock should have the highest absolute correlation with the proxy proposed by \cite{romer2004new}. The upper part of figure \ref{fig:proxyandLoadings} plots the posterior distribution of the correlation of the shocks with the proxy series. Again, we find clear evidence that the first shock has the highest correlation with the proxy, while the correlation of the other shocks with the proxy is rather low. Finally, we look at the posterior distributions of the shocks at specific dates. This allows us to examine whether if they are consistent with economic narratives, see \cite{antolin2018narrative}. \cite{antolin2018narrative} argue that the monetary policy shock was positive (contractionary) for the observations corresponding to April 1974, October 1979, December 1988 and February 1994, and negative for December 1990, October 1998, April 2001, and November 2002. In addition, \cite{antolin2018narrative} argues that in October 1979 a major contractionary monetary policy shock greatly increased the fed funds rate. In figure \ref{fig:Narratives} we plot the posterior distribution of the monetary policy shocks times the corresponding factor loadings from the fed funds rate equation for the eight time points. Remember that all values of the posterior distribution of the the factor loadings are positive, see figure \ref{fig:proxyandLoadings}. We find that the posterior distributions of the first shock have the correct sign for all eight dates. Moreover, we also find that the first shock was the main driver of an unexpected increase in the fed funds rate in October 1979. These results further strengthen the interpretation of the first shock as a monetary policy shock.

To label a monetary policy shock, we have used economic reasoning that could also have been used to identify a monetary policy shock by relying only on the second moments of the data. In this case, however, we have to impose this information as restrictions that cannot verified with the data. By contrast, by exploiting information in higher moments of the data we do not need to impose economic restrictions but can instead confirm economic reasoning.

\subsection{Checking the Identifying Assumptions }
\begin{figure}[t]
	\centering
	\includegraphics[width=1\linewidth]{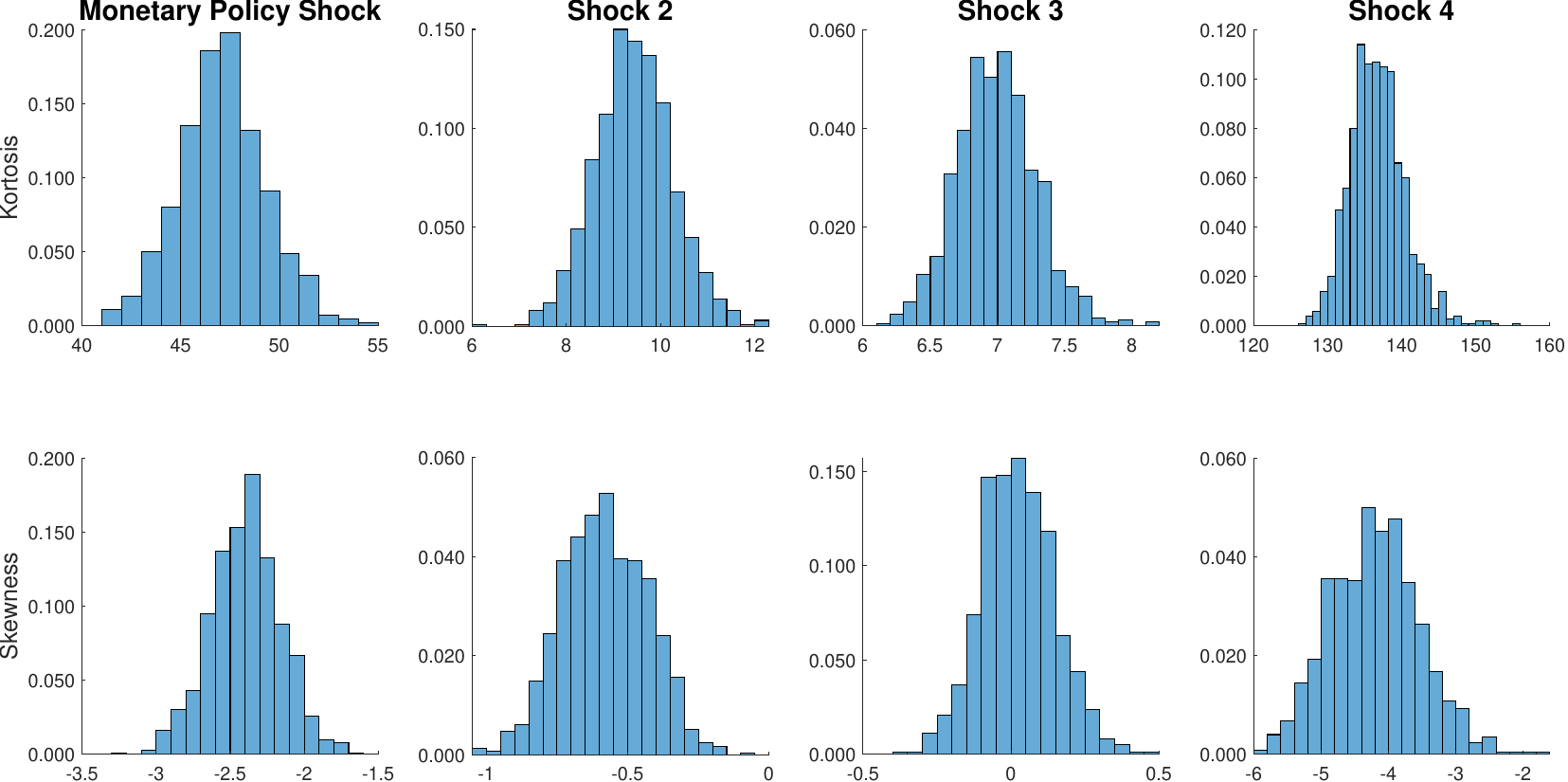}
	\caption{The first panel shows the posterior distributions of the shocks kurtosis. The second panel shows the posterior distributions of the shocks skewness.}
	\label{fig:NonGaussian}
\end{figure}

\begin{figure}[t]
	\centering
	\includegraphics[width=1\linewidth]{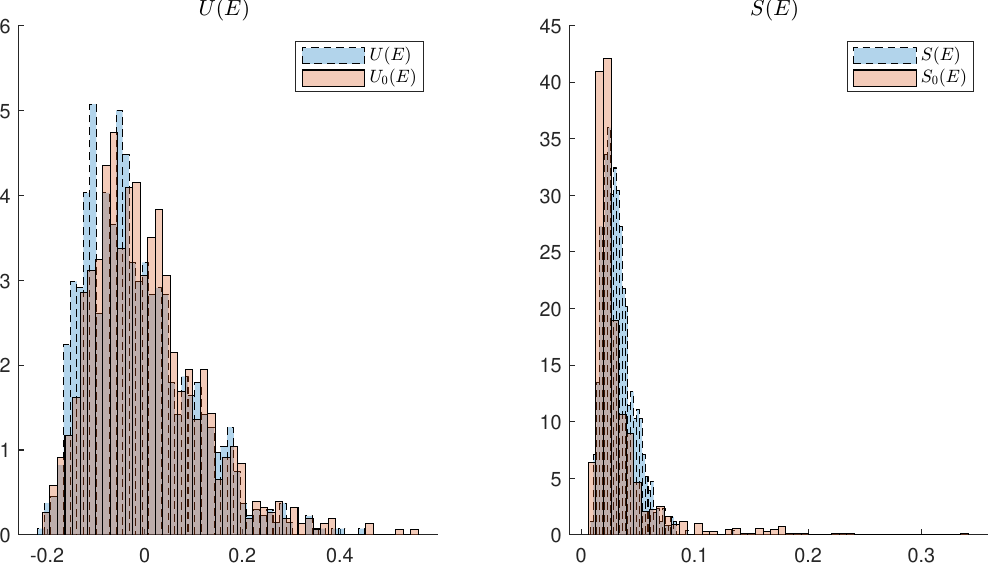}
	\caption{The left panel plots the posterior distribution of the test statistic in \cite{matteson2017independent} $(U(E))$ as well as the posterior distribution of repermuted shocks $(U_0(E))$. The right panel plots the posterior distribution of the test statistic in \cite{montiel2022svar} $(S(E))$ as well as the posterior distribution of repermuted shocks $(S_0(E))$.   }
	\label{fig:Independence}
\end{figure}

It is useful to assess the empirical plausibility of assuming non-Gaussian and mutually independent structural shocks.  Figure \ref{fig:NonGaussian} shows the posterior distributions of the skewness and kurtosis of the structural shocks. For all shocks, we find a sizeable degree of non-Gaussianity in the structural shocks. In particular, the kurtosis has positive values far above three. The monetary policy shock distribution is left skewed, which indicates that large negative Fed surprises tend to be larger than large positive fed surprises in an absolute sense.
Nest, we look at the plausibility of the mutual independence assumption. We follow \cite{braun2023importance} and report posterior distributions of popular frequentist test statistics. The first is a nonparametric test developed in \cite{matteson2017independent}. Let $E=(\bm f_1,\dots,\bm f_T)'$ denote the $T\times r$ structural shocks. The statistic is given by $U(E)=T\sum_{j=1}^{K-1} \mathcal{I}_T(\hat{U}_k,\hat{U}_{j+})$, where $j+=\{l:j<l\leq K\}$ denotes the indices $(j+1,\dots,K)$, $\hat{U}_j$ has elements defined as $\hat{u}_{i,k}=\frac{1}{T} \text{rank} \{f_{ij}: f_{ij}\in E_j\}$, and $\mathcal{I}_T$ is the empirical distance covariance as defined in \cite{matteson2017independent}. While this test is consistent against all types of dependence, others may have higher power against certain alternatives. \cite{montiel2022svar} propose an alternative testing for shared volatility in structural shocks. They consider the test statistic $S(E)=\sqrt{\frac{1}{K(K-1)}\sum_{i=1}^K\sum_{j\neq i}\text{Corr}(f_{it}^2,f_{j,t}^2)^2}$, which measures the root of the mean squared sample cross-correlations of squared structural shocks. Figure \ref{fig:Independence} shows the posterior of these two test statistics. As in \cite{braun2023importance}, we overlap each distribution with that of the same statistic computed for randomly repermuted socks, denoted by $U_0(E)$ and $S_0(E)$. This helps to get an indication of how the posterior of the test  statistic would look like under the null of mutual independence. Both distributions $U(E)$ and $S(E)$ largely overlap with the distributions based on resampled shocks, suggesting no evidence against mutual independence.

\subsection{Impulse Response Functions}
\begin{figure}[t]
	\centering
	\includegraphics[width=1\linewidth]{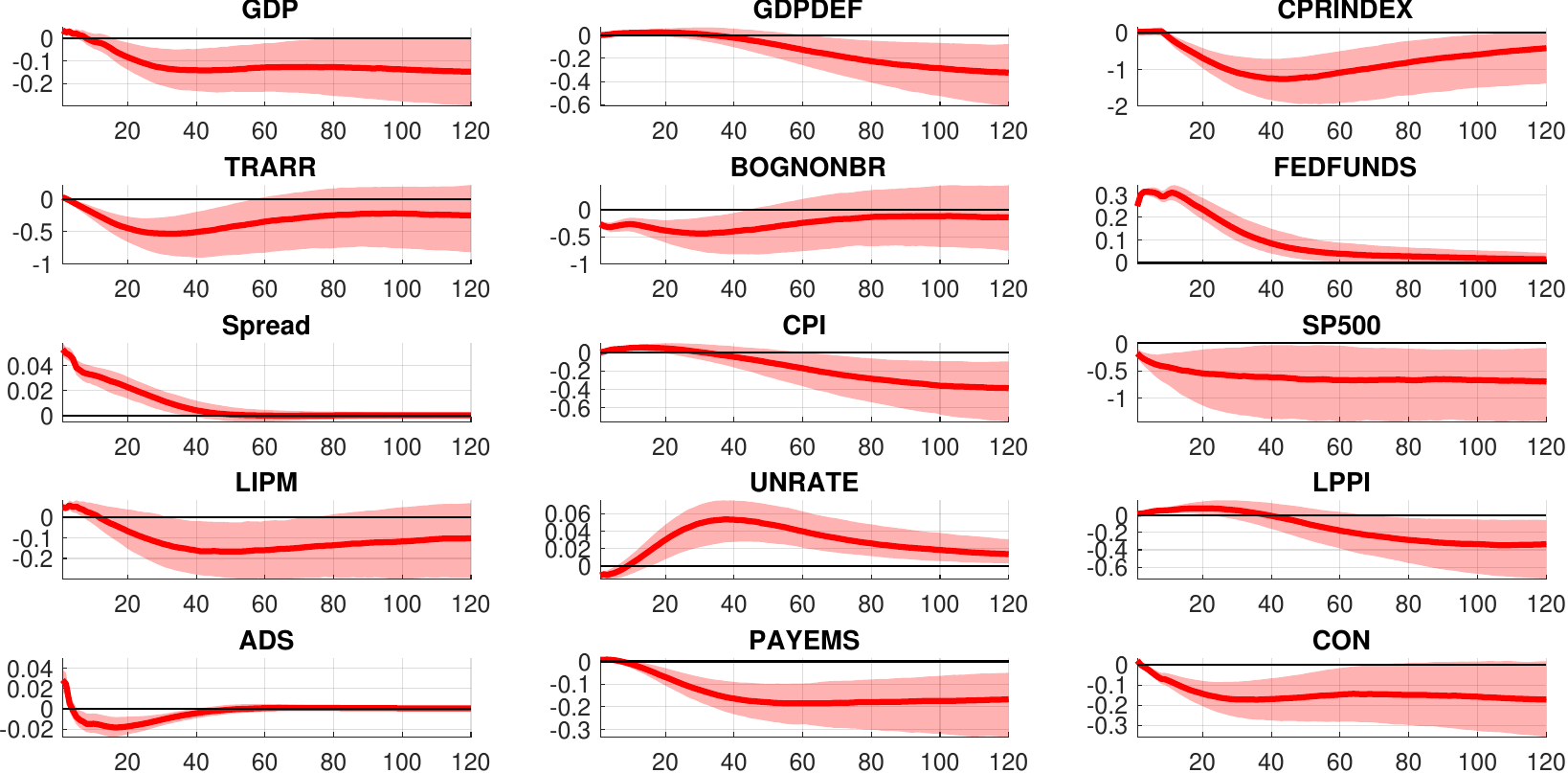}
	\caption{The figure show the responses to a monetary policy shock. }
	\label{fig:IRFs}
\end{figure}

We now turn to the impulse responses to the identified monetary policy shock. 
Figure \ref{fig:IRFs} shows the median impulse responses along with their 68\% credible bands.
As the model is estimated using standardised data,
the IRFs are standardised back such that the magnitude can be interpreted in the unit of
measurement with respect to table \ref{tab:Data}. We normalise the Feds funds rate by 0.25 basis points for the impact period.

The response of the real GDP to the monetary policy shock, which was the subject of Uhlig (2005), is slightly
positive in the first periods and then becomes persistently negative. The marked delay in the transmission of the monetary policy shock to output and the persistently negative response are in line with standard economic intuition. However, it is in contrast to \cite{uhlig2005effects}, who find a positive effect of a contractionary monetary policy shock on output. 

In line with \cite{uhlig2005effects} and \cite{antolin2018narrative}, we find the effect of a positive (contractionary) monetary policy shock on the commodity price index and
central bank reserves to be both negative and persistent. In contrast, we find that the response
of the GDP deflator (and other price measures) to be slightly positive or zero (at least in the short run), which may simply indicate a significant delay in the transmission of monetary policy to the deflator, as is the case for real GDP. Note however, that \cite{uhlig2005effects} and \cite{antolin2018narrative} restrict the GDP deflator to be negative in the first six months.

Our results are very consistent with those of the seminal paper by \cite{romer2004new}. In particuar, our estimated response of real GDP to a contractionary monetary policy shock is remarkably similar to theirs, despite being derived solely from information provided by higher moments, without relying on their detailed quantitative or narrative records. Moreover, consistent with our results, \cite{romer2004new} also find a significant delay to two years in the transmission of monetary policy shocks to prices.

As mentioned earlier, there are usually several data series corresponding to the same
economic variable. And it is often unclear which of these should be used, if only one variable
is to be selected. For example, in our application, the time series GDP deflator, consumer price index and
producer price index are all good candidates for the economic variable prices. Similarly, we use real GDP and industrial production to measure economic activity and use unemployment and employment as proxies for the labour market.

The median responses of the real GDP and industrial production are negative and have very similar shapes. However, their credible intervals are somewhat different. The credible bands of industrial production are wider than those of real GDP. If ony real GDP had been used, a stronger conclusion might have been drawn than it is justified. Similarly, the responses of the various price variables have very similar shapes. However, the produce price index suggests a slightly larger initial increase in prices than the GDP deflator. 

Comparing the response of the unemployment rate with that of employment, we find that the unemployment response mirrors the response of real GDP (initially falling until rising persistently) while the employment response is delayed until becoming persistently negative. In contrast to the output and labour market variables, we find that real consumption starts to fall immediately after the shock period. This again highlights the benefit of including more variables and conducting a more comprehensive structural analysis.
Plausibly, the financial variables respond without any delay. The response of the spread is positive and the response of stock prices is negative, in accordance with economic theory. Overall, we find that output and prices respond with a large delay to the monetary policy shock.

\subsection{Extending the Model with a Proxy Variable}

In this section, we combine identification by higher moments with identification by proxy variable as discussed in section 2.5. We use the proxy variable suggested by \cite{romer2004new}. To assess the empirical plausibility the proxy being exogenous, we estimate two versions of the model. The first version imposes the proxy restriction that only the target shock is allowed to be correlated with the proxy variable. This restriction is imposed by placing zero restriction on the matrix of factor loadings $\bm L$, see section 2.5. The second version is estimated without these zero restrictions. For both versions we compute the DIC reported in table \ref{tab_DIC}. The DIC for the model without the zero restrictions is lower than for the model with the zero restrictions, providing evidence against the zero restrictions. Thus, we provide empirical evidence  against exogenous exclusion restrictions. This result is consistent with \cite{braun2023identification}. 

\begin{table}[H]
	\centering
		\captionof{table}{Deviance Information Criteria for Proxy restrictions}
		     \begin{threeparttable}
	\begin{tabular}{cc}
		\toprule
	\textbf{Proxy restrictions} &\textbf{No restrictions}  \\
		\midrule
       -141151 & -213888 \\
		\bottomrule
	\end{tabular}
 \begin{tablenotes}\footnotesize
    \item[]
    Notes: The table contains the DIC of the model with proxy zero restrictions and without. Small values are preferred.
    \end{tablenotes}
		 \end{threeparttable}
	\label{tab_DIC}
\end{table}
\vspace{-10pt}

\section{Conclusion}
In this paper we propose a large structural VAR with a factor structure. Non-Gaussian and mutually  independent factors provide statistically identification of the matrix of factor loadings without the need to impose economically motivated restrictions. These factors are interpreted as structural shocks. Attaching an economic meaning to the statistically identified shocks allows as to perform structural analysis in a large dimensional setting. We propose a Gibbs sampler to estimate the model and develop an estimator of the DIC. The DIC can be used to decide between different model specifications. Finally, we discuss how economic restrictions can be added to the model. We highlight the benefit of the model using both artificial as well as real data. Experiments with artificial data show that our model possesses good estimation properties. In the empirical application, we show how we can identify a monetary policy shock and provide empirical evidence that prices and output respond with a large delay to a monetary policy shock.

%
%
%
%
%
%

\small{\setstretch{0.85}
\addcontentsline{toc}{section}{References}
\bibliographystyle{frbcle.bst}
\bibliography{Lit}
\appendix

\renewcommand{\thesection}{Appendix \Alph{section}}
\setcounter{table}{0}
\renewcommand{\thetable}{\Alph{section}.\arabic{table}}
\setcounter{figure}{0}
\renewcommand{\thefigure}{\Alph{section}.\arabic{figure}}


\section{Data}

\begin{table}[ht]
	\centering
	\begin{tabular}{llll}
		\toprule
	\textbf{Abbreviation} &\textbf{ Variable}&\textbf{transformation}&\textbf{Source} \\
		\midrule		
		GDP &Real gross domestic product&log times 100& Uhlig \\
		GDPDEF &GDP deflator&log times 100& Uhlig \\
		CPRINDEX &Commodity price index&log times 100& Uhlig \\
		TRARR &Total reserves&log times 100&Uhlig \\
		BOGNONBR &Non-borrowed reserves&log times 100&Uhlig\\
		FEDFUNDS &Federal funds rate&none& Uhlig \\
		Spread & Commercial paper spread &none&Uhlig \\
		CPI &Consumer Price Index&log times 100& Uhlig \\
		SP500 &S\&P500 index&log times 100& Uhlig \\
		LIPM &Manufacturing industrial production&log times 100&FREDMD \\
				UNRATE &Unemployment rate&none&FREDMD \\
		LPPI &Producer price index&log times 100&FREDMD \\
		ADS &Business condition index&log times 100&FREDMD \\
		PAYEMS &All Employees: Total nonfarm&log times 100&FREDMD \\
						CON &Real personal consumption expenditures&log times 100&FREDMD \\
		\bottomrule
	\end{tabular}
	\caption{The tables shows the data used in the empirical application as well as their transformations, sources and abbreviations. Time series with "`Uhlig"' were obtained from the replication files of \cite{arias2019systematic}.
Note that GDP and GDPDEF were interpolated based on US industrial production and CPI prices, respectively. The Commercial paper spread is calculated as 3-month AA financial commercial paper rate minus the 3-months T-bill rate. Time series with "`FREDMD"' are obtained from the dataset of \cite{mccracken2016fred}.}
	\label{tab:Data}
\end{table}

\end{document}